\documentclass[review]{elsarticle}

\usepackage{hyperref}
\usepackage{dingbat}
\usepackage{xcolor}
\definecolor{darkblue}{rgb}{0, 0, 0.5}
\hypersetup{colorlinks=true,citecolor=darkblue, linkcolor=darkblue, urlcolor=darkblue}
\usepackage{float}
\usepackage{verbatim} 
\usepackage{apalike}
\usepackage{soul}
\restylefloat{figure}
\restylefloat{table}

\usepackage{adjustbox}
\usepackage{graphicx}%
\usepackage{multirow}
\usepackage{amsmath,amssymb,amsfonts}%
\usepackage{amsthm}%
\usepackage{mathrsfs}%
\usepackage[title]{appendix}%
\usepackage{xcolor}%
\usepackage{textcomp}%
\usepackage{manyfoot}%
\usepackage{booktabs}%
\usepackage{algorithm}%
\usepackage{algorithmicx}%
\usepackage{algpseudocode}%
\usepackage{listings}%
\usepackage{natbib}
\usepackage{multirow}
\usepackage{enumitem}
\newlist{steps}{enumerate}{1}
\setlist[steps, 1]{label = \arabic*:}
\usepackage{tabularray}
\usepackage{booktabs}
\usepackage{multirow}
\usepackage{caption}
\usepackage{subcaption}
\usepackage{xcolor}
\definecolor{_yellow}{HTML}{FFF2CC}
\definecolor{_blue}{HTML}{DAE8FC}
\usepackage[T1]{fontenc}

\usepackage[utf8]{inputenc}
\usepackage{etoolbox}
\AfterEndEnvironment{enumerate}{\vskip-\lastskip}

\DeclareMathOperator*{\argmax}{arg\,max}

\DeclareMathOperator*{\meanpool}{Mean\,Pool}

\journal{Computer Speech and Language}

\biboptions{authoryear}

\begin{document}
\begin{frontmatter}

\begin{titlepage}
\begin{center}
\vspace*{1cm}

\textbf{ \large Enhancing Event Extraction from Short Stories through Contextualized Prompts}

Chaitanya Kirti$^{a}$ (ckirti@iitg.ac.in), Ayon Chattopadhyay$^{b}$ (ayon.chattopadhyay@iitg.ac.in), Ashish Anand$^{b,a}$ (anand.ashish@iitg.ac.in), Prithwijit Guha$^{c,a}$ (pguha@iitg.ac.in) \\

\hspace{10pt}

\begin{flushleft}
\small  
$^a$ Centre for Linguistic Science and Technology, Indian Institute of Technology Guwahati, Assam, India \\
$^b$ Department of Computer Science and Engineering, Indian Institute of Technology Guwahati, Assam, India \\
$^c$ Department of Electronics and Electrical Engineering, Indian Institute of Technology Guwahati, Assam, India

\vspace{1cm}
\textbf{Corresponding Author:} \\
Chaitanya Kirti \\
Centre for Linguistic Science and Technology, Indian Institute of Technology Guwahati, Assam, India \\
Tel: (+91) 9431680757 \\
Email: ckirti@iitg.ac.in

\end{flushleft}        
\end{center}
\end{titlepage}

\title{Enhancing Event Extraction from Short Stories through Contextualized Prompts}

\author[label1]{Chaitanya Kirti\corref{cor1}}
\ead{ckirti@iitg.ac.in}

\author[label2]{Ayon Chattopadhyay}
\ead{ayon.chattopadhyay@iitg.ac.in}

\author[label2,label1]{Ashish Anand}
\ead{anand.ashish@iitg.ac.in}

\author[label3,label1]{Prithwijit Guha}
\ead{pguha@iitg.ac.in}

\cortext[cor1]{Corresponding author.}
\address[label1]{Centre for Linguistic Science and Technology, Indian Institute of Technology Guwahati, Assam, India}
\address[label2]{Department of Computer Science and Engineering, Indian Institute of Technology Guwahati, Assam, India}
\address[label3]{Department of Electronics and Electrical Engineering, Indian Institute of Technology Guwahati, Assam, India}

\begin{abstract}
Event extraction is an important natural language processing (NLP) task of identifying events in an unstructured text. 
Although a plethora of works deal with event extraction from new articles, clinical text etc., only a few works focus on event extraction from literary content.
Detecting events in short stories presents several challenges to current systems, encompassing a different distribution of events as compared to other domains and the portrayal of diverse emotional conditions. This paper presents \texttt{Vrittanta-EN}, a collection of 1000 English short stories annotated for real events. Exploring this field could result in the creation of techniques and resources that support literary scholars in improving their effectiveness. This could simultaneously influence the field of Natural Language Processing. Our objective is to clarify the intricate idea of events in the context of short stories. Towards the objective, we collected 1,000 short stories written mostly for children in the Indian context. Further, we present fresh guidelines for annotating event mentions and their categories, organized into \textit{seven distinct classes}. The classes are {\tt{COGNITIVE-MENTAL-STATE(CMS), COMMUNICATION(COM), CONFLICT(CON),  GENERAL-ACTIVITY(GA), LIFE-EVE-\\NT(LE), MOVEMENT(MOV), and OTHERS(OTH)}}. Subsequently, we apply these gu-\\idelines to annotate the short story dataset. Later, we apply the baseline methods for automatically detecting and categorizing events.
We also propose a prompt-based method for event detection and classification. The proposed method outperforms the baselines, while having significant improvement of more than 4\% for the class \texttt{CONFLICT} in event classification task. 

\end{abstract}


\end{frontmatter}

\section{Introduction}
\label{sec1}

\begin{figure}[h]
    \centering
    \includegraphics[scale=0.8]{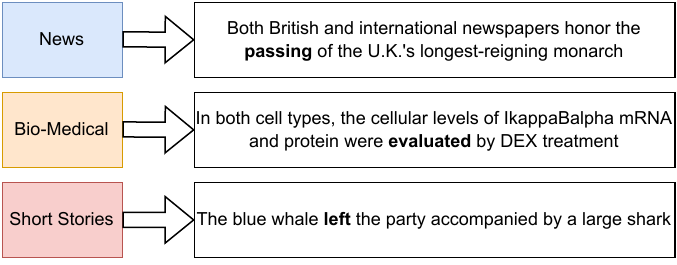}
    \caption{Examples of event triggers in News, bio-medical and short stories. News and Biomedical domains generally deal with real-life scenarios, while short stories may contain real-life scenarios in unreal situations.}
    \label{fig:example}
\end{figure}

The process of extracting structured information from a text, such as entities, relations, events, etc., is known as information extraction (IE). Event extraction, a crucial component of IE, is responsible for locating trigger mentions of a specific event type and those events' participants. An event is a particular occurrence at a particular time and place, involves one or more individuals, and is often defined as a change of states \citep{eventdefACL}. 
The task of event extraction comprises three subtasks: 
(1) Event detection (ED) that pulls trigger words from the text to indicate when an event occurs. Trigger words are those words that most clearly express the occurrence of an event; (2) Assigning a predefined event type to event triggers; and (3)  Event argument extraction, i.e., to identify an event's participants. The task is challenging because there are many ways to convey the same event, and same phrase can express distinct events. For example, \textit{left} could mean a \texttt{movement}, as in \autoref{fig:example}, or could mean the act of not covering the food, as in ``the food has been left uncovered''.

Event extraction is useful for information retrieval~\citep{basile2014extending}, question answering~\citep{yang2003structured}, narrative schema extraction~\citep{chambers2009unsupervised}, reading comprehension and summarization~\citep{cheng-erk-2018-implicit}, and for building knowledge banks (KBs).  Event extraction has several applications in various domains. Structured events, for example, can be used explicitly to extend knowledge bases, allowing for more logical reasoning and inference\citep{eearea1}. Governments have long focused their public affairs management on event identification and tracking, as understanding the outbursts and evolution of popular social events allows them to respond quickly \citep{eearea2}. In the financial domain, event extraction can assist businesses in rapidly identifying consumer reactions to their goods and deriving indications for risk analysis and trading recommendations \citep{eearea3}. Event extraction can be used in the biomedical domain to identify the alterations in biomolecules such as genes and proteins or interactions between two or more biomolecules \citep{eearea4}. In short, advances in event extraction techniques and systems can support a wide range of domains.

Previous works primarily cover datasets in the domain of news, bio-medical etc., involving neural methods for event extraction~\citep{sims2019literary}. Event extraction from short stories and the classification of those events remains a less explored area. This motivated us to create a dataset of Indian short stories. Our objective is to make a valuable contribution to resource building and support future research in this domain through the introduction of the \texttt{Vrittanta-EN} corpus.

As prompt-based generative methodologies have recently gained prominence across all NLP tasks including for the event extraction, we thoroughly assess the quality of \texttt{Vrittanta-EN} through both (1) neural approaches, and (2) the prompt-based methods. We adhered to prior research in prompt-based generative event extraction models utilizing PLMs with an encoder-decoder transformer architecture, similar to models like BART \citep{lewis-etal-2020-bart}, and T5 \citep{JMLR:v21:20-074}. Our proposed methodology employs the BART model as its PLM, which is a conventional Transformer-based PLM featuring both an Encoder and a Decoder. 
These evaluations provide a comprehensive understanding of the dataset's reliability and effectiveness, enabling us to validate its suitability for a wide range of applications and research endeavours. Detailed annotation guidelines, annotated dataset, and top-performing models are accessible on GitHub.
 


The importance of gold-standard data in building Natural Language Processing (NLP) applications cannot be overstated. As NLP strives to handle more intricate linguistic phenomena and diverse labelling tasks, the demand for substantial labelled datasets remains high~\citep{sun2017revisiting, human-annotation}. However, the annotation process has become more challenging and costly due to these complexities. Manually annotating linguistic resources for training and testing, particularly in specialized text domains, is a resource-intensive and expensive endeavour. So, to expand the dataset, the best-performing model was employed, whose quality was established experimentally. 

The main contributions of this work are as follows:
\begin{itemize}
\item Introduces a novel annotation guideline for event trigger detection and event classification tailored for Indian short stories with a specific focus on capturing real events, i.e., events that reflect actual occurrences.
\item Presentes \texttt{Vrittanta-EN}, the first annotated corpus of 200 short stories using the above guideline. Initially, 200 short stories were manually annotated. Subsequently, the dataset is expanded to 1000 through the trained model and human feedback.
\item Extensive experiments to establish essential baselines for the dataset along with  ablation studies.
\item Proposes a prompt-based span extraction approach for the event extraction and classification task, thereby mitigating the problem of data imbalance.
\end{itemize}
\vspace{0.3 cm}

The rest of the paper is organized as follows. Section~\ref{section:prev} presents previous work. Section~\ref{section:dataset} presents the dataset construction procedure, followed by the description of EVENT PROMPT model in Section~\ref{section:prompts}. \ref{section:exp} presents the experiments conducted. Section~\ref{section:res} describes the results and discussions. The errors are analysed in Section~\ref{section:error}. Finally, Section~\ref{section:conc} concludes the paper.


\section{Previous Work}
\label{section:prev}
Several public evaluation programs (PEPs) have been initiated to enhance the competitiveness of event extraction, including MUC~\citep{sundheim1993survey}, TDT~\citep{allan2002topic}, ACE~\citep{ace}, TAC-KBP~\citep{mcnamee2010evaluation}, and SemEval\footnote{\url{https://aclanthology.org/venues/semeval/}}. They have provided standardized evaluation resources, encouraged competition among researchers, and set benchmarks for performance measurement.




The primary requirement of any event extraction task is detecting specific types of events mentioned in the annotated data. In the existing literature, the event extraction task has been divided into open and closed domains \citep{xiang2019survey}. Closed-domain event extraction uses a predefined event schema to discover and extract desired events of a particular type from text \citep{hogenboom2011overview}. ACE 2005 \citep{doddington2004automatic}, ERE \citep{song2015light} and TAC-KBP \cite{mitamura2015overview} use closed domain method by defining the structure and type of events first and then extract results based on the defined structure. The standard dataset ACE 2005 corpus includes 8 types of events, with 33 subtypes~\citep{ace}. 

The lack of annotated corpus of biological literature led to the development of the GENIA corpus \citep{kim2003genia}. It allows NLP techniques to work for text mining at the biomolecular level by providing semantically annotated reference scientific articles. 
Another corpus in the clinical domain for tracking the disease condition, observing treatment outcomes and complications, and detecting side effects in medication was introduced to the research community by the organizers of the i2b2 Challenge \citep{sun2013evaluating}. 
Another corpus builds on top of TimeBank, adding to it semantic information for the annotation dealing with the factuality of events in text \citep{sauri2009factbank}. A corpus of sentences is annotated as general or specific by annotators on Amazon Mechanical Turk from news articles comprising reports on events, finance, and science journalism \citep{louis2012corpus}. DBpedia \citep{knuth2015dbpedia} is another dataset created based on live extraction of Wikipedia. The authors follow rule sets to detect events, such as news related to sports, politics, real life, etc., from any update in Wikipedia content.
The recognition and elaboration of historical events is a crucial step when dealing with historical texts. 
A historical corpus named Histo Corpus \citep{sprugnoli2019novel} contains historical text from travel narratives and news, published between the late nineteenth and early twentieth century. 

The role of events in literary fiction, however, is very different from their role in fact-based reporting of events in the real world, including historical texts~\citep{sprugnoli2019novel}. In literature, imaginary causalities exist, which are very difficult to extract in contrast to the hard-coded causalities in fact-based reports~\citep{sauri2009factbank}. Novels tend to be much longer than news articles~\citep{louis2012corpus} or clinical documents~\citep{nedellec2013overview} and tend to have more complex narrative structures. Since event extraction from historical texts, clinical documents, or news has now become possible due to the availability of datasets, the same was not possible for the literature domain. However, recently, ~\cite{sims2019literary} have annotated a new dataset of literary events, Litbank\footnote{\url{https://github.com/dbamman/litbank/tree/master/events}}. This dataset focuses solely on events in literary texts that are depicted as happening actually. It has taken literary texts from novels published in the nineteenth and twentieth centuries. In terms of techniques used, existing closed-domain approaches can be divided into four categories: pattern matching \citep{riloff1993automatically}, machine learning \citep{liao2010using}, deep learning \citep{zheng2019doc2edag}, and prompt-based approaches.

Many pattern-based event extraction systems have been created for many application areas, including biomedical event extraction \citep{cohen2009high}, \citep{casillas2011using} and financial event extraction \citep{borsje2010semi}, as a result of the AutoSlog system extracting terrorist incidents~\citep{riloff1993automatically}.
To extract biological events, Casillas et al. \citep{casillas2011using} used Kybots (Knowledge Yielding Robots). To manually generate event patterns, the Kybots system uses a rule-based approach.
 Cao et al. \citep{cao2018including} proposed that further event patterns can be generated by integrating the ACE corpus with other expert-defined patterns.

In the machine learning approach, a nearest-neighbor learning method is used for recognizing triggers and a maximum entropy learner(s) classifier is adopted for identifying arguments~\citep{ahn2006stages}. Event patterns and token attributes can also be combined as input for machine learning classifiers \citep{meyers2001parsing}.
In learning deep or complex nonlinear relations, deep learning-based methods can overcome the limits of classic machine learning methods. 
A CNN has been shown to be effective at capturing the syntactic and semantics of a sentence \citep{kalchbrenner2014convolutional} because it is capable of learning text-hidden features based on continuous and generalized word embeddings. Nguyen and Grishman \citep{liu2017improving} may be the first to build a CNN for event detection, that is, detecting a trigger and the sort of event it causes in a sentence. As the network input, each word is first turned into a real-valued vector representation that is a concatenation of the word embedding, its location embedding, and the entity type embedding. The CNN consists of a convolution layer, a max pooling layer, and a softmax layer from input to output, and outputs the classification result for each word.
The attention mechanism has recently become popular in several NLP tasks \citep{liu2017exploiting}. \cite{zeng2016convolution} recommended that each word's local contextual representation be learned first with a CNN, then concatenated with the output of another Bi-LSTM to generate the final word representation for classification. The generative adversarial network (GAN) is another important type of hybrid model, which typically comprises of two neural networks competing against each other~\citep{hong2018self}.

Prompt-based learning methods, separately treated as a single category, model various sequence-to-sequence tasks with the help of pre-trained Encoder-Decoder-based Transformer architectures like BART \citep{lewis-etal-2020-bart}, and T5 \citep{JMLR:v21:20-074}.
The prompts help the model understand the specific requirements of the task and improve its performance by providing the model with relevant context for understanding and extracting events from text. These prompts typically consist of a combination of text and placeholders that guide the model on what information to extract and how to structure the output.
In contrast to the pre-training and fine-tuning paradigm, which requires a lot of resources and data, prompt-based techniques transform the downstream tasks into a form, more aligned with the model's pre-training tasks. By creating linked prompts with blanks and determining a mapping from certain filled words to projected categories,~\cite{schick-schutze-2021-exploiting} converts a number of classification tasks into cloze activities. They construct related prompts and try to map the blanks, that have been used for constructing the prompts, to predicted categories. \cite{li-liang-2021-prefix} mainly work on generation tasks and propose a lightweight prefix-tuning approach where they freeze the model parameters and only adjust a sequence of continuous task-specific vectors. 
The task in  BART \citep{lewis-etal-2020-bart} is extended for event argument extraction, where the events are required to be given as the input to predict its arguments~\citep{ma2022prompt, zhong2023contextualized, peng2024prompt}. These models focus on event argument extraction, leaving the tasks of event detection and classification unaddressed. Therefore, there is a gap where the subtasks of event extraction and classification can be completed using prompt-based learning methods. 
To put it differently, there was almost negligible effort to discover a definition of an event that pertains to the domain of short stories. This contrasts with the approach taken in other fields, such as the news or clinical domain. Moreover, the potential of contextualized prompts has not been leveraged. This serves as the primary reason driving the present study.

\section{Dataset Construction}
\label{section:dataset}
This section covers the source of the short stories, key annotation guidelines, and dataset statistics. Furthermore, detailed statistics about the event classes used during the annotation process are presented. 
This extends our previous work, which focused solely on event detection with single-word event triggers~\citep{kirti}. In this extension, we explore event classification alongside event detection, expanding our scope to include multi-word event expressions.

\subsection{Source of the short stories }

The dataset encompasses a collection of 1000 short stories obtained from the public domain. These stories are centred around the Indian context and primarily cater to children. 
The stories are drawn from a variety of interconnected fable collections like:
\begin{itemize}
    \item \textbf{Panchtantra (PT)} - 
Fables featuring animals, characterized by their light, vibrant nature, are suitable even for very young children. They offer valuable lessons that leave a lasting impression on young minds.
    \item \textbf{Champak (CP)} - These tales, blending humans and animals, carry messages of self-assurance, compassion, and perseverance through elements of entertainment, mystery, exploration, and scientific discovery.
    \item \textbf{Tenali-Raman (TR)} -  These entertaining tales for children illustrate how cleverness, intellect, and sagacity triumph over sheer force and dominance.
    \item \textbf{Akbar-Birbal (AB)} - These narratives recount the dynamics between emperor Akbar and his counsellor, Birbal. Combining humour with valuable life lessons provides entertainment and moral enlightenment.
    \item \textbf{Betal Pachisi (BP)} - a collection of tales and legends within a frame story, from India. 
    \item \textbf{Hitopadesh (HP)} -  a collection of tales featuring a mix of animals and humans, conveying maxims, practical wisdom, and political counsel in straightforward, refined language.
    \item \textbf{Jataka Tales (JT)} - A compilation of stories and moral tales portraying various incarnations of Siddhartha Gautama, the future Buddha, depicted alternately as animals or humans.
    \item \textbf{Singhasan Battisi (SB)} - a collection of Indian folk tales narrated by the 32 lions adorning King Vikramaditya's legendary throne, emphasizing virtues, wisdom, and valor.  
\end{itemize}
 
From these eight sources, 25 stories were selected from each for annotation. This rich assortment of stories serves the crucial purpose of delving into a wide range of events and exploring various writing styles and narrative techniques. Each event within these stories is accompanied by an event trigger.
These event triggers, typically expressed as verbs, although occasionally as nouns and also adjectives, are specifically designed to elicit the occurrence of that particular event. 




\subsection{Annotation Guideline}

In this subsection, the key annotation guideline that was utilized for annotation is discussed. Short stories possess unique characteristics that set them apart from other types of texts. Detecting events within short stories poses a specific challenge, especially considering their target audience of children and the narrative structures employed. In these stories, it is common to encounter animals or even inanimate objects engaging in conversations with one another. The occurrence of events in such stories relies heavily on the narrative, genre, and storytelling style. Considering the prior research findings, we have created fresh annotation guidelines centered on recognizing and categorizing event mentions to address these specific challenges for events in short stories. The issues and rules addressed by the guidelines established in ACE~\cite{ace}, NewsReader~\cite{eventdefNRG}, HISTO Corpus \cite{sprugnoli2019novel}, and Litbank \cite{lit} are taken into account. By leveraging insights from prior research, the aim is to enhance the accuracy and effectiveness of event detection within short stories. The proposed guidelines are valuable resources for annotators, as clear instructions and criteria are provided for identifying and labeling events within the context of children's short stories. Hereafter, bold words represent realis events, and underlined word shows non-real events that seem to be misunderstood as real events.

\subsubsection{Event Expression}
Our annotation framework focuses on realis events in short stories. It considers the following linguistic components that could represent an event. Each case is illustrated with at least one example.
\begin{itemize}
    \item \textbf{Finite and non-finite verbs}:
    \begin{itemize}
    \item[(1)]He \textbf{painted} the fence.
    \item[(2)]He \textbf{wanted} to play.
    \end{itemize}
\end{itemize}

\begin{itemize}
    \item \textbf{Past participles}: In nominal pre-modifier position representing the resultative event and as a sentential predicate.
    \begin{itemize}
    \item[(3)]A \textbf{frustated} man.
    \item[(4)]Her father is \textbf{retired}.
    \end{itemize}
\end{itemize}

\begin{itemize}
    \item \textbf{Present participles}: In the nominal pre-modifier position representing in-progress events.
    \begin{itemize}
    \item[(5)]Children are \textbf{playing} in the garden.
    \end{itemize}
\end{itemize}
\begin{itemize}
    \item \textbf{Nouns}:
    \begin{itemize}
    \item[(6)]His third \textbf{marriage} was with Uma Devi.

\noindent Pronouns are not tagged as an event. There are some cases when the noun refers to the participants (noun), and the same refers to the occurrence of the event. In those particular cases, it is not tagged.    
    \item[(7)]The \underline{retired} were \textbf{going} for lunch.
    
    \noindent\textbf{Stand-Alone Noun:}\\When there are more than two triggers possible for a single event, a noun is tagged if that noun is referring to the event.

    \item[(8)] They \underline{launched} an \textbf{attack}.
    \end{itemize}
\end{itemize}

\begin{itemize}
    \item \textbf{Adjectives};\\ \noindent\textbf{Stand-Alone Adjective:} Whenever an adjective and a verb are used together, the adjective is selected if that adjective is referring to the event.
    \begin{itemize}
    \item[(9)]They \underline{looked} \textbf{terrified}.
    \end{itemize}
\end{itemize}

\noindent\textbf{Complex Examples:}
\begin{enumerate}
    \item There is a possibility that a single event is triggered by more than a single word.
    \begin{itemize}
        \item The parties \underline{held} a \textbf{meeting} in Delhi.
    \end{itemize}
        Both held and meeting can be tagged as triggers but with the help of the stand-alone noun rule, we choose meeting as the trigger.
    
    \item There is also a possibility that multiple events are present in the same scope.
    \begin{itemize}
            \item The \textbf{attack} \textbf{injured} 15 and \textbf{killed} 3.
    \end{itemize}
        Here 3 events are going to be tagged: \textbf{attack} as {\tt{CONFLICT}}  and \textbf{injured} and \textbf{killed} as {\tt{LIFE-EVENT}}.
        
    \item Rules for differentiating between cases whether two triggers refer to the same event or different events.
    \begin{enumerate}
        \item If a person doing one event is the same person doing the other, there is a single event.
        \item If one event is a subpart of another event, then they are the same event.
        \item If one event is describing another event’s internal structure, they are the same, else they are two different events.
        \item In all other confusing cases, we always consider presence of  two different events. 
            \begin{itemize}
                \item Gandhiji was \textbf{shot dead} by Godse.
                Here, we tag two separate events. Shot is tagged as a {\tt{CONFLICT}} event, and dead is tagged as {\tt{LIFE-EVENT}}.
                \item The hurricane \underline{left} 20 \textbf{dead}.\\
                In this example, only \textbf{dead} would be annotated because left and dead are used to express the same event.
            \end{itemize} 
    \end{enumerate}
        
\end{enumerate}

\subsubsection{Dimensions to capture realis events} The dimensions followed in the Light ERE approach~\citep{aguilar2014comparison} are considered to capture realis events and are mentioned below.   
\begin{enumerate}
    \item\textbf{Polarity:} Events must have a positive polarity and negative polarity is ignored. 
\begin{itemize}
    \item[(13)] Ram \textbf{gave} the test and didn’t \underline{fail}.
    \item[(14)] Ganesha had not \underline{moved} from his spot.
    \item[(15)] Bheema did not even \underline{look up}.
\end{itemize}
\item\textbf{Tense:} Events must be in the past or present tense. Events in the future tense are not tagged.

\begin{itemize}
      \item[(14)] He \textbf{traveled} to Delhi the last September.
    \item[(15)] They will \underline{meet} next month.
\end{itemize}

\item\textbf{Generality:} Only specific events are tagged; all generic events are ignored. Specific events are events with singular occurrences at a particular place and time.
\begin{itemize}
 \item[(16)] I \textbf{ate} rice for dinner last night. 
    \item[(17)] People often \underline{eat} rice for dinner.
\end{itemize}

\item\textbf{Modality:} Actual occurrences are tagged. All other modalities (believed, hypothetical, desired, etc.) are not.
\begin{itemize}
 \item[(18)] He \textbf{walked} to \underline{buy} some cold drinks.
 \item[(19)] They would not \underline{trouble} the Devas again.
 \item[(20)] Rumors of \underline{arrests} circulated in Mumbai.
  \item[(21)] If you keep \underline{grabbing} a dozen people every day and \underline{eating} them up, soon there would be no more people \underline{left} in the town.\\
\end{itemize}
\end{enumerate}

\subsubsection{Event Classes} Every event that has been annotated needs to be categorized to be a particular class. Event categorization is required to know the distribution of different events present in the domain of short stories. As the events in short stories diverge significantly from earlier datasets, adhering to the categorization outlined in prior literature was opted against. The events are classified into seven types so as to cover all the major happenings. The classes are {\tt{COGNITIVE-MENTAL-STATE (CMS), COMMUNICATION (COM), CONFLICT (CON),  GENERAL-ACTIVITY (GA), LIFE-EVENT (LE), MOVEMENT (MOV), and OTHERS (OTH).}}\\


\noindent {\tt{I. COGNITIVE-MENTAL-STATE:}} This event type describes mental states and mental acts that involve mental or cognitive processes e.g. think, know, remember, perceive, prefer, want, forget, understand, decide, decision, excitement/calmness, pleasure/suffering, compassion/indifference, courage/fear, love/ hate emotional actions, states and processes, lack of emotions, goodness or badness, inferiority or importance.
    \begin{itemize}
        \itemindent=5pt
        \item[(22)] The witnesses were \textbf{amazed} at the man’s calmness.
        \item[(23)] Abhimanyu, while in his mother’s womb, had \textbf{learned} how to break the Chakravyuh.
        \item[(24)] Lord Brahma \textbf{decided} to test Krishna’s powers.
        \item[(25)] He was \textbf{shocked} to see all the children and calves 
    \end{itemize}
         Emotional actions like hugging, smiling, laughing, smiling, etc are tagged in this type.

    \begin{itemize}
        \itemindent=5pt
        \item[(26)] Piku \textbf{smiled} at Riku.
        \item[(27)] Raju started \textbf{laughing} after hearing the joke.
    \end{itemize}

\noindent {\tt{II. COMMUNICATION:}} When the event shows that the communication happened between two entities, it is tagged as {\tt{COMMUNICATION}}. The subtypes of the event have been categorized such as:-

\begin{enumerate}
 
    \item \textbf{Physical Communication:} Like two or more people
talking or having a conversation physically in front of each other, e.g. say, talk, ask, etc.
\begin{itemize}
     \item[(28)] They were \textbf{asked} to aim at the eye of a toy bird on the branch of a tree. 
 \end{itemize}

    \item\textbf{Virtual Communication:} Interaction of people with each other without being in the same room. Use of letters, messages, emails, phone calls, etc.
    \begin{itemize}
    \item[(29)] Rohan \textbf{sent} an email to Priya.
    \item[(30)] Students joined the virtual \textbf{meeting}.
    \end{itemize}
    Here, ``meeting'' is considered a trigger because of the stand-alone noun rule.

    \item\textbf{Announcement/Declaration:} Announcement through media or press, an organization declaring something, naming things, producing speech acts, e.g. report, announcement, called.
    \begin{itemize}
        \item[(31)] The next day the principal made an \textbf{announcement}.
        \item[(32)] It was an \textbf{announcement} about a cooking contest in the colony.
    \end{itemize}

    \item We also tag event that shows the spreading of news or things vocally as {\tt{COMMUNICATION}}.\\
\end{enumerate}

\noindent {\tt{III. CONFLICT:}} An event is classified as {\tt{CONFLICT}} if it shows any kind of conflict, disturbance, fight (both verbally or non-verbally), or argument. The subtypes of {\tt{CONFLICT}} are-
\begin{enumerate}
    \item \textbf{Attack:} Fight started by one party
    \begin{itemize}
       \item[(33)] U.S. forces continued to \textbf{bomb} Fallujah.
       \end{itemize}

    \item \textbf{Fight-War:} Fight between 2 entities
    \begin{itemize}
       \item[(34)] On the thirteenth day of the great \textbf{battle.}
       \item[(35)] The sixteen-year-old youth \textbf{fought} bravely against the experienced.
    \end{itemize}

    \item \textbf{Demonstrate:} Whenever a large number of people come together in public to protest or demand some sort of official action.
    \begin{itemize}
       \item[(36)] The union began its \textbf{strike} on Monday.\\
    \end{itemize}
       
\end{enumerate}

\noindent {\tt{IV. GENERAL-ACTIVITY: }} In {\tt{GENERAL-ACTIVITY}} all general daily life events are tagged. The domain of general activities is very large. The subtypes more clearly express which kind of general events are to be tagged.

\begin{enumerate}
\item \textbf{Physical-Meet:} Face-to-face meeting, identified individual.
\begin{itemize}
      \item[(37)] Rohan and Sibu \textbf{met} this week.\\
Here, ``met'' should not be confused with the {\tt{COMMUNICATION}} event type. A virtual meeting is considered part of communication because it calls/meets someone virtually to talk to, whereas a physical meeting is like meeting someone and not talking, or it can happen occasionally, or it may be a non-planned activity.

\end{itemize}

\item \textbf{General:} It covers general activities that happen in life such as doing, using, trying, helping, success, failure, having, losing, taking, giving, sharing, seeing, watching, sleeping, wake-up, carrying, eating drinking, hearing, standing, dancing, singing, performance, saving, owning, wet-dry, etc.  
\begin{itemize}
    \item[(38)] The best tobacco was \textbf{used} in the cigar factory.

    \item[(39)] A man is \textbf{trying} to free himself with the help of a band.
\end{itemize}

\item \textbf{Animal:} Animal-related things and activities like biting, sound, roar, etc.
\begin{itemize}
    \item[(40)] The dog \textbf{barked} at the thief.\\
\end{itemize}
\end{enumerate}

\noindent {\tt{V. LIFE-EVENT:}} Life event are mostly inspired by the ACE(2005)\citep{ace}. These subtypes explore the event tags by {\tt{LIFE-EVENT}}.
\begin{enumerate}
\item\textbf{Birth:} Events related with birth.
\begin{itemize}
\item[(41)] Devaki and Vasudeva’s eighth child Krishna was \textbf{born} at the stroke of midnight.
\end{itemize}

\item\textbf{Injure:} Events related to some physical harm.
\begin{itemize}
\item Rahul was \textbf{wounded} in the accident.
\item Taga was so badly \textbf{hurt} that he had to be taken to a hospital in an ambulance.
\end{itemize}

\item\textbf{Medical Assistance:} Medical treatment, physical pain, diseases, diagnosis, operation, etc.
\begin{itemize}
    \item Raju’s leg \textbf{operation} is going well.
\end{itemize}

\item\textbf{Die:} This is related to end of life. Some examples of trigger words are die, kill, murder, assassinate, late, suicide, etc.
\begin{itemize}
\item Abhimanyu \textbf{killed} many great warriors before he himself was \textbf{killed}. 
\end{itemize}

\item\textbf{Live:} Person Is living. 
\begin{itemize}
\item Two tribes \textbf{lived} on the banks of a river.
 
\end{itemize}

\item\textbf{Plant:} Plant and forest-related things and activities.
\begin{itemize}
   \item Oranges are \textbf{growing}. 
\end{itemize}

\item\textbf{Marry:} Two people are married under the legal definition.
\begin{itemize}
\item He'd been \textbf{married} before and had a child.
\item Rahul and Riya got \textbf{engaged} yesterday. 

\end{itemize}

\item\textbf{Divorce:} Two people are officially divorced under the legal definition of divorce; this type also considers the separation of couples.
\begin{itemize}
    \item The couple \textbf{divorced} four years later.\\
\end{itemize}

\end{enumerate}

\noindent {\tt{VI. MOVEMENT:}} When any kind of movement happens, regardless of the medium, such as air, water, or hard surface.
\begin{enumerate}
\item\textbf{Travel:} When a single entity or group of entities are moving or traveling somewhere.
\begin{itemize}
    \item Krishna \textbf{crawled} to the two trees in his courtyard.
    \item Ram \textbf{went} to Kiskindha.
\end{itemize}

\item\textbf{Transport:} When something or someone is transported through some vehicle or medium.
\begin{itemize}
    \item The weapons were \textbf{moved} to a secure site.
    \item The company \textbf{sent} the parcel in the morning. \\
\end{itemize}
\end{enumerate}

\noindent {\tt{VII. OTHERS:}} Events that do not fit within the aforementioned six categories will be categorized under the {\tt{OTHERS}} category. This category is regarded as an outlier category, and the rationale for its inclusion is to encompass all possible remaining events.





\subsection{Dataset Statistics}
The proposed annotation guidelines adapts the problems and rules addressed by ACE~\citep{ace} for the chosen data. Only the realis events, i.e., the events that are actually happening, with their event classes are considered ~\citep{doddington2004automatic}.
For capturing realis events, four aspects - Polarity, Tense, Genericity, and Modality have been selected~\citep{aguilar-etal-2014-comparison};~\citep{song2015light};~\citep{lit}.

\begin{table*}
\centering
\scalebox{0.60}{\resizebox{\textwidth}{!}{
\begin{tabular}{lllll}
\hline
\textbf{Sl.No} & \textbf{Statistics} & \textbf{Count} \\
\hline
1. & Total tokens in the dataset & 174,559  \\
2. & Total unique tokens in the dataset & 10,484  \\
3. &  Total sentences in the dataset & 10,259 \\ 
4. & Average tokens per story & 873  \\ 
5. &Average sentences per story & 51  \\
6. &  Average tokens per sentence & 17   \\ 
7. &  Total events in the dataset & 11,272 \\
8. &  Average events per story & 56 \\ 
\hline
\end{tabular}}}
\caption{Statistics of the annotated dataset (200 short stories)}
\label{tab:raw_data}
\end{table*}

We annotate 200 stories based on the proposed guideline. An equal number of 25 stories have been considered from each of the sources to avoid biases related to a particular narrative. The web-based BRAT annotation tool~\citep{a18} is used for annotation.
The number of words in the stories ranges between 500 to 1500. Two annotators make all the annotations based on the proposed guideline.
This study presents an annotated dataset consisting of realis events from Indian short stories. \autoref{tab:raw_data} shows a detailed breakdown of the dataset statistics. The dataset contains 174,559 tokens, with a significant vocabulary diversity indicated by 10,484 unique tokens. Each story averages 873 tokens and 51 sentences, suggesting moderately lengthy narratives. With an average of 17 tokens per sentence, the stories exhibit concise sentence structures. The dataset records 11,272 events, averaging 56 events per story, indicating a high event density. This suggests that the stories are rich in eventful content, providing a comprehensive resource for studying narrative structures and event annotation.

\begin{table*}
\centering
\scalebox{0.75}{\resizebox{\textwidth}{!}{
\begin{tabular}{lllll}
\hline
 \textbf{Event Types} & \textbf{Total Count} & \textbf{Train} & \textbf{Dev} & \textbf{Test}  \\
\hline
{\tt{COMMUNICATION}}           &    3456 &    2052 &    447 &    957 \\     
{\tt{GENERAL-ACTIVITY}}     &    2951 &    1837 &    325 &    789  \\       
{\tt{MOVEMENT}} &    2192 &    1329 &    294 &    569    \\       
{\tt{COGNITIVE-MENTAL-STATE}}               &    1528 &    849 &    178 &    501  \\    
{\tt{LIFE-EVENT}}               &    535 &    314 &    60 &    161   \\ 
{\tt{OTHERS}}             &    490 &    295 &    55 &    140  \\      
{\tt{CONFLICT}}               &    120 &    62 &    30 &    28  \\ 
\hline
{Total}               &    11,272 &    6,738 &    1,389 &    3,145  \\  
\hline
\end{tabular}
}
}
\caption{Number of different events present in the annotated dataset}
\label{tab:event_typ}
\end{table*}

\begin{figure}[h!]
    \centering
    \includegraphics[scale=0.36]{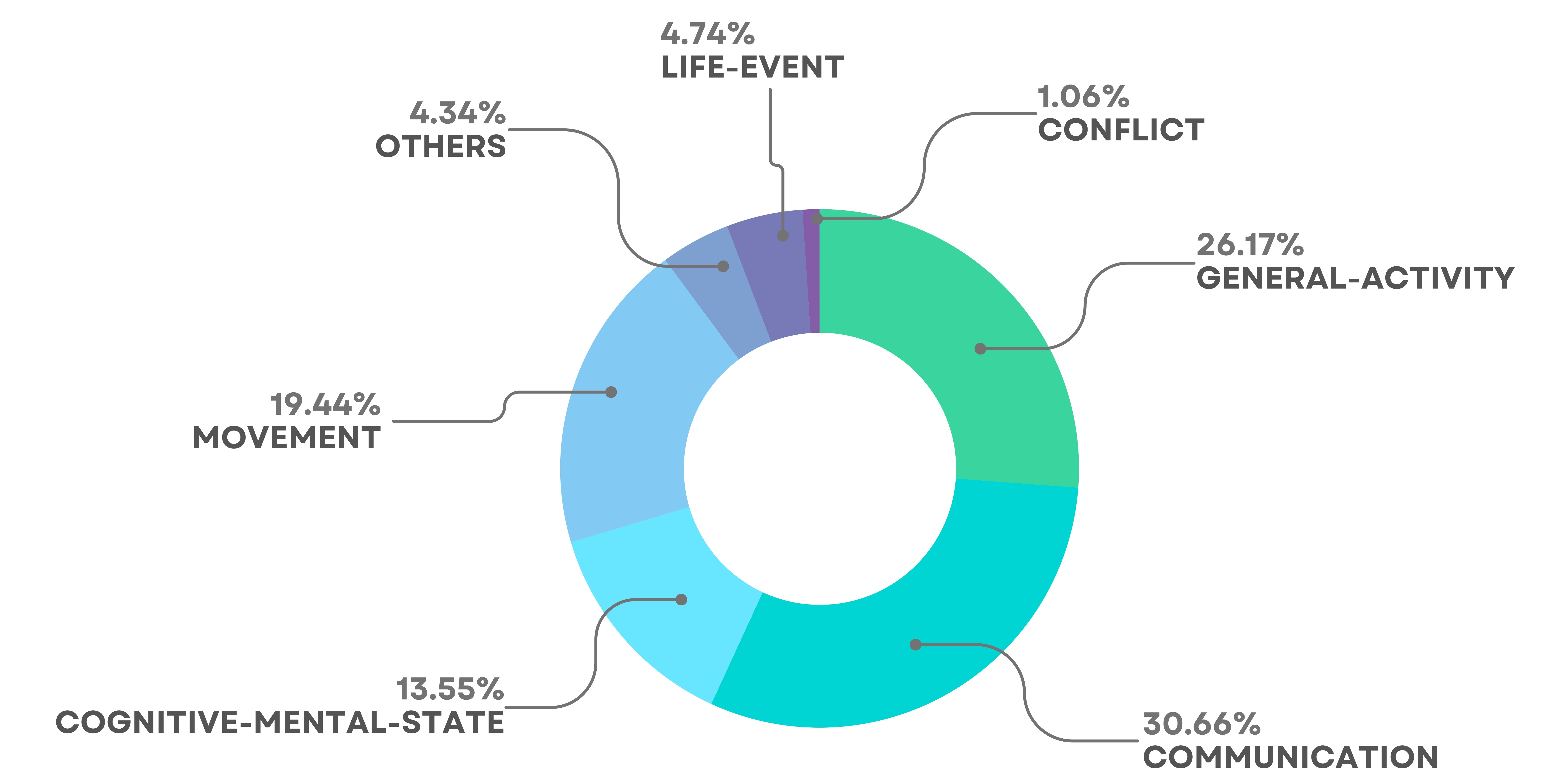}
    \caption{Distribution of different events in the dataset}
    \label{fig:pie}
\end{figure}

\begin{figure}[h!]
    \centering
    \includegraphics[scale=0.65]{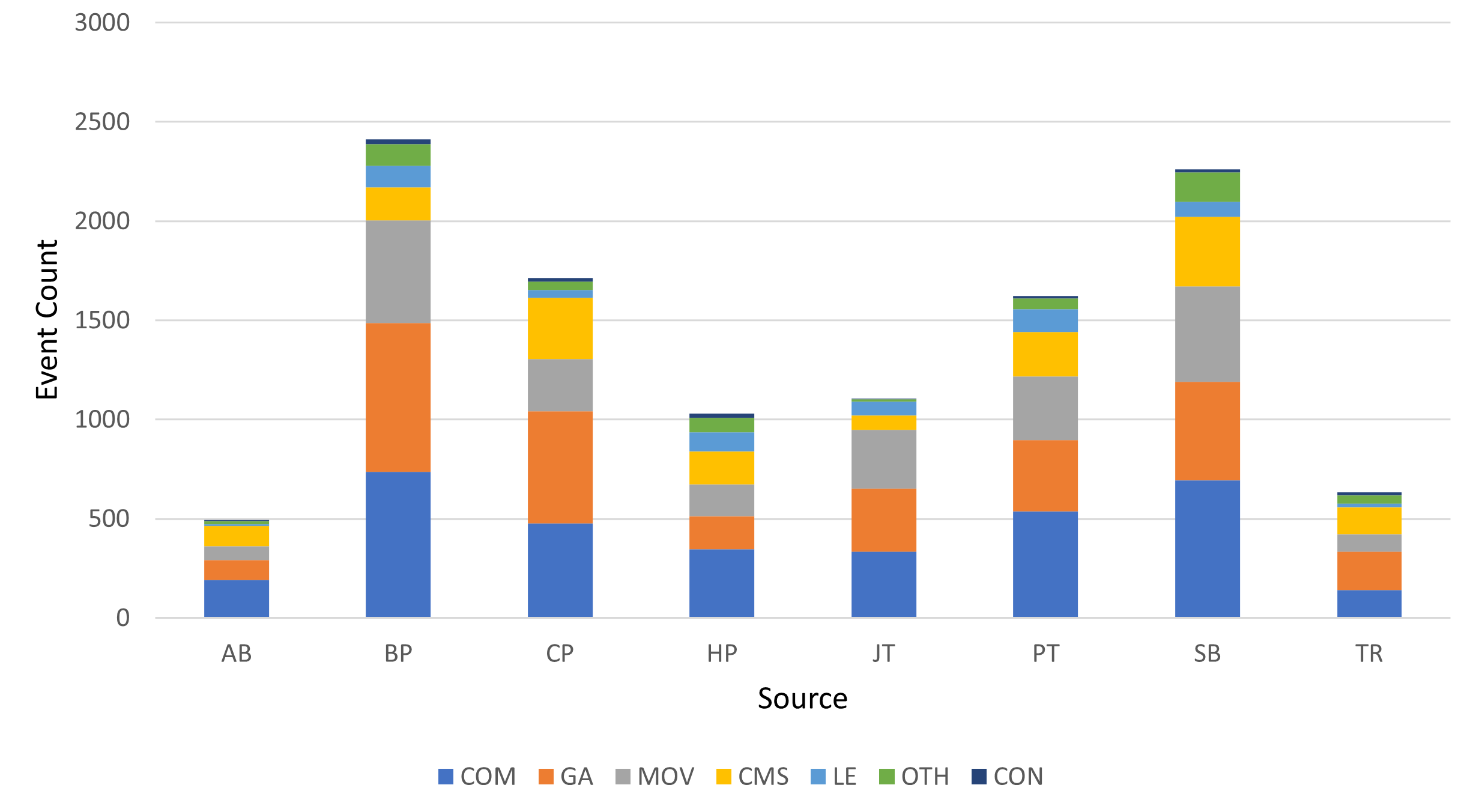}
    \caption{Distribution of events from different sources in the dataset}
    \label{fig:bar}
\end{figure}

\begin{figure}[!t] 
    \centering
    \subfloat[un-annotated corpus]{%
        \includegraphics[width=0.50\linewidth]{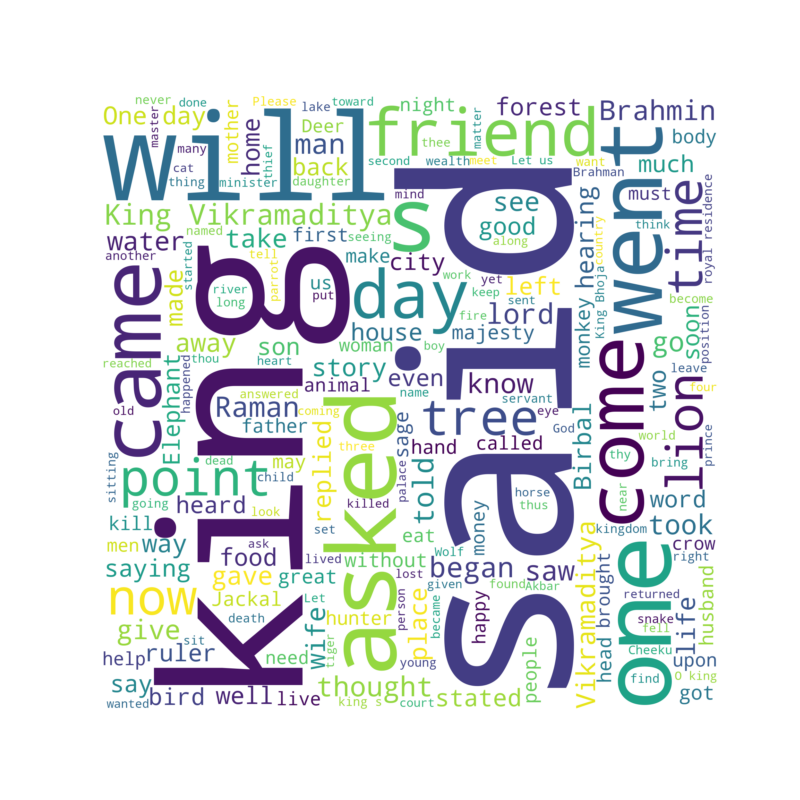}%
        \label{prompt1}%
        }%
    \hfill%
    \subfloat[only trigger words]{%
        \includegraphics[width=0.50\linewidth]{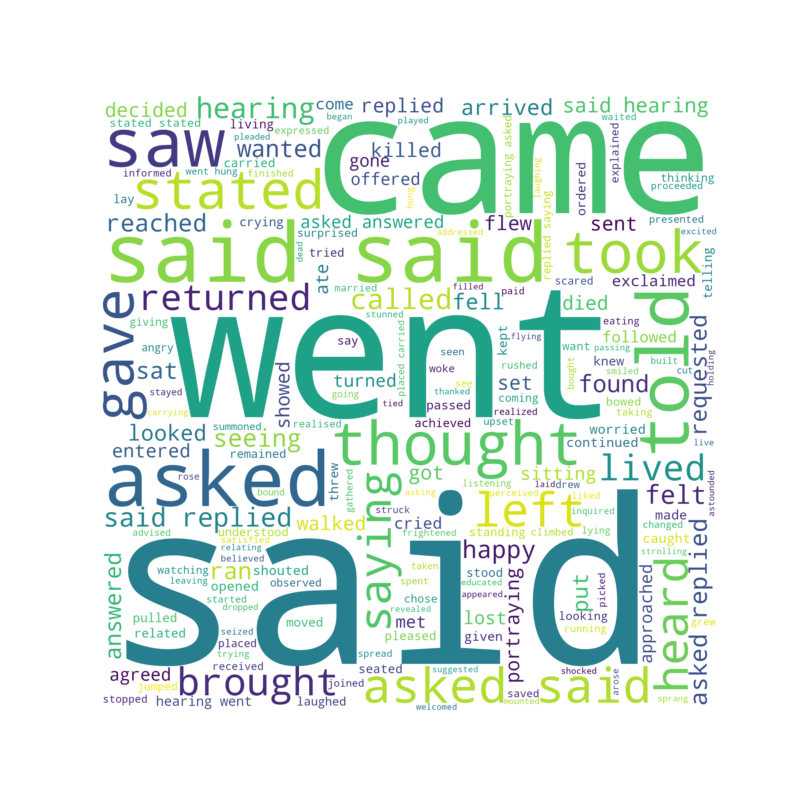}%
        \label{prompt2}%
        }%
    \caption{Word cloud of un-annotated corpus and trigger words of the entire dataset}
    \label{fig:wordcloud}
\end{figure}

\autoref{tab:event_typ} categorizes the number of events in the train, development (dev), and test sets across seven event types within the dataset. The total number of events is 11,272, distributed with 6,738 events in the train set, 1,389 in the dev set, and 3,145 in the test set. The most frequent event type is {\tt{COMMUNICATION}}, with 3,456 occurrences, followed by {\tt{GENERAL-ACTIVITY}} with 2,951 occurrences. {\tt{MOVEMENT}} and {\tt{COGNITIVE-MENTAL-STATE}} are moderately common, with 2,192 and 1,528 events, respectively. {\tt{LIFE-EVENT}}, {\tt{OTHERS}}, and {\tt{CONFLICT}} are less frequent, indicating they are rarer in the dataset. The distribution across the train, dev, and test sets is consistent, ensuring a balanced representation of each event type, which is crucial for robust model training and evaluation. This detailed annotation provides a rich resource for analyzing and modelling event-based narratives.
\autoref{fig:pie} represents the distribution of the events in the dataset. The dataset clearly shows that {\tt{COMMUNICATION}} events are plentiful (30.66\%), whereas {\tt{CONFLICT}} events are rare (1.06\%). The distribution of the events from different short stories is shown in~\autoref{fig:bar}. 
\autoref{tab:ann_data} represents the top 15 words tagged as an event in the annotated dataset. \autoref{fig:wordcloud} presents the word cloud of the entire dataset and for trigger words only. It is clear that trigger words with the highest frequencies are in the majority of the entire dataset. This can also be correlated with \autoref{tab:ann_data}. Words like ``said'', ``asked'', ``went'', ``came'', ``told'', etc., are evident in the dataset. 

Two annotators, who are co-authors of this work, performed the annotation task. 20 random stories were chosen, and both the annotators annotated the stories in accordance with the provided annotation guidelines. An inter-annotator agreement (IAA) of 83.1\% was achieved. The IAA serves as a meaningful indicator of the clarity of the annotation guidelines, the consistency in the annotators' comprehension of these guidelines, and the overall feasibility of the annotation task. The remaining 180 stories were annotated by a single annotator, resulting in a total of 200 annotated stories.

\begin{table*}
\centering
\scalebox{0.5}{\resizebox{
\textwidth}{!}{\begin{tabular}{lll}
\hline
\textbf{Trigger Word} & \textbf{Count} &\textbf{Event Rate}\\
\hline
said &	1298 & 96\%\\
asked&	415 & 98\%\\
went	&409 & 95\% \\
came	&291 & 82\%\\
told &	190 & 92\%\\
saw	&179 & 92\% \\
replied &	170 & 98\%\\
thought&	148 &81\%\\
took&	142 & 78\%\\
hearing&	113 & 92\%\\
saying	&96 & 67\%\\
heard	&92 & 74\%\\
brought	&81 & 86\%\\
returned &76 & 94\%\\
answered	&75 & 91\%\\
\hline

\end{tabular}}}
\caption{Top 15 words that are annotated as event triggers. Its count and event rate are also shown. Event rate is the percentage of labelled events with respect to the total occurrence of that word in the corpus}
\label{tab:ann_data}
\end{table*}

\subsection{Evaluation Metrices}
We used the identical evaluation criteria as previous studies~\citep{wadden2019entity}, which presented the Precision (P), Recall (R), and F1 scores for event detection. We defined TP as a true positive, FN as a false negative, and FP as a false positive. 
\[P=\frac{TP}{TP+FP}\quad\]
\[R=\frac{TP}{TP+FN}\quad\]
\[F1=2*\frac{P*R}{P+R}\quad\]

True Positives are the cases where the model predicted a positive outcome (e.g., a class label) correctly, and the ground truth label is also positive. False Positives are the cases where the model predicted a positive outcome, but the ground truth label is actually negative. In other words, the model made a false positive prediction. False Negatives are the cases where the model predicted a negative outcome, but the ground truth label is actually positive. In other words, the model failed to detect a positive instance, resulting in a false negative prediction.
The F1 score provides a better indication of the model's event detection performance, as it considers both the recall rate and precision.

\section{Contextualized Prompts} 
\label{section:prompts}
Prompt-based learning represents a novel approach within the domain of pre-trained language models~\citep{liu2023pre}. Diverging from the traditional pre-training and fine-tuning paradigm, prompt-based techniques transform downstream tasks into formats that align more closely with the model's pre-training objectives. Contextualized prompts can be used in conjunction with pre-trained language models to fine-tune the model for event extraction tasks. 

\subsection{Problem Formulation}
We formulate the problem of event extraction as a prompt-based span extraction task. Prompt here stands as an input token sequence that provides us with information we want to extract from the context. Considering $\mathcal{D} = [D_1, D_2, ..., D_n]$ as the entire dataset and $E = [e_1, e_2, ..., e_l]$ as the set of event types. Each document $D_i = [X_1, X_2, ..., X_m]$ where $X_j$ is the context/sentence that is given as input to our architecture. Our aim is to extract a span $a^{(e_k)}$ $\forall e_k \in E$. If $X_j$ does not have any trigger $t$ for the event type $e_k$ then the span detected should be $(0, 0)$, i.e., an invalid span.

\subsection{Proposed Architecture}

\begin{figure}[ht]
    \centering
    \includegraphics[scale=0.58]{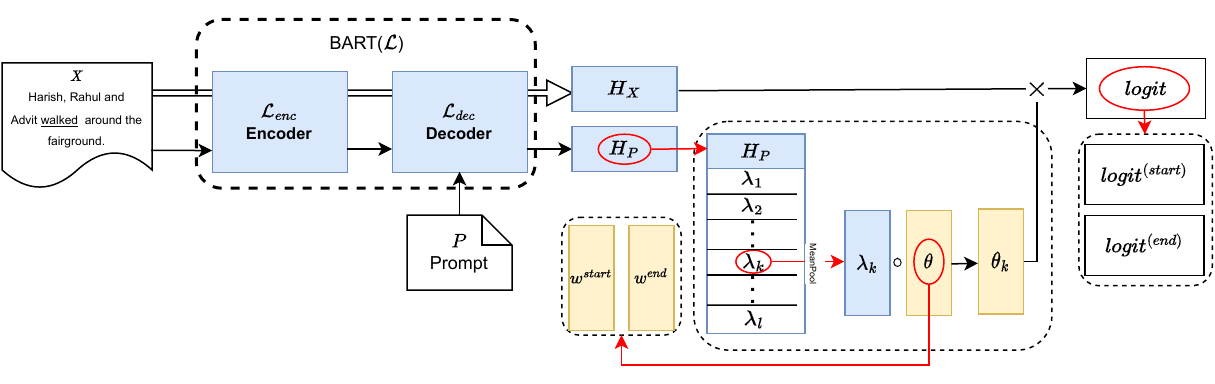}
    \caption[Prompt-Based Architecture]{Representation of the prompt-based architecture for event extraction and classification. $\circ$ denotes element-wise multiplication, $\times$ denotes matrix multiplication. Underlined tokens are predicted when the corresponding input is being used.\colorbox{_yellow}{Yellow} represents that weights are updated in the layer while \colorbox{_blue}{Blue} represents that weights are not updated in the layer.}
    \label{fig:piee}
\end{figure}

The proposed architecture is depicted in the \autoref{fig:piee}. Here, we aim to extract event triggers and their spans. The classification of these event triggers is also done.  

\subsubsection{Prompt Creation}
We create a single prompt for the types of events that we want to extract from the dataset $\mathcal{D}$. The presence of event types in the prompt is referred to as the \textit{slots} for the events. We don't allow multiple events of the same type to occur in a context, because the dataset $\mathcal{D}$ doesn't have multiple events of the same type in the same context. In our work, we have used concatenation to create the prompt where we concatenate all the event types that we want to extract from $\mathcal{D}$ as shown in the \autoref{tab:prompts}. 



\begin{table}[]
\begin{adjustbox}{width=\columnwidth,center}
\caption{Prompts that are used by the prompt creator function}
\label{tab:prompts}
\begin{tabular}{|llll|}
\hline
\multicolumn{4}{|c|}{\textbf{PROMPTS}}                                                                                                                     \\ \hline
\multicolumn{1}{|l|}{EVENT DETECTION}      & \multicolumn{3}{l|}{EVENT}                                                                                    \\ \hline
\multicolumn{1}{|l|}{EVENT CLASSIFICATION} & \multicolumn{3}{l|}{CONFLICT COMMUNICATION LIFE-EVENT MOVEMENT COGNIIVE-MENTAL-STATE GENERAL-ACTIVITY OTHERS} \\ \hline
\end{tabular}
\end{adjustbox}
\end{table}

\subsubsection{Learning prompt-based span selection}
Considering the context $X \in \mathbb{R}^{t_x \times d_e}$ where $t_x$ is the length of input tokens and $d_e$ is the output dimension of feature embedding, along with prompt $P$ of length $t_p$ as the conditioning input, we first aim to create a learnable span selector for each event $e_k \in E$.
We choose BART~\citep{lewis-etal-2020-bart}, a pre-trained language model built on a standard transformer that combines both an \textit{encoder} and a \textit{decoder} which we represent here as $\mathcal{L} = [\mathcal{L}_{enc}, \mathcal{L}_{dec}]$. We do not concatenate the prompt $P$ and the context $X$, instead, we give the context $X$ to the BART-Encoder $\mathcal{L}_{enc}$ and prompt $P$ to BART-Decoder $\mathcal{L}_{dec}$. $X$ and $P$ would interact mutually in the cross-attention layers of $\mathcal{L}_{dec}$.

\begin{gather}
    H^{(enc)}_X = \mathcal{L}_{enc}(X) \in \mathbb{R}^{t_x \times d_{enc}} \\
    H_X = \mathcal{L}_{dec}(H^{(enc)}_X ; H^{(enc)}_X) \in \mathbb{R}^{t_x \times h}\\
    H_P = \mathcal{L}_{dec}(P ; H^{(enc)}_X) \in \mathbb{R}^{t_p \times h}
\end{gather}
where $H_X$ denotes the context representation and $H_P$ denotes to context-oriented prompt representation. If we want to extract the context-oriented prompt representation for the $k^{th}$ event we have to mean-pool its corresponding representation in $H_P$ and obtain the event feature $\lambda_k \in \mathbb{R}^h$ where $h$ denotes the dimension of the last hidden layer of $\mathcal{L}_{dec}$. 
\begin{gather}
    H_{P_k} = ({H_P}_{ij}) ~ \forall_j \in [P^{st}_k, P^{ed}_k] \mid H_{P_k} \in \mathbb{R}^{end(t_{p_k}) - start(t_{p_k}) \times h}\\
    \lambda_k = \meanpool(H_{P_k}) \in \mathbb{R}^{h}\\
\end{gather}
where $P^{st}_k$ is start and $P^{ed}_k$ is the end of $k^{th}$ slot in prompt $P$. After extracting the representation for the $k^{th}$ event, we can utilize the same for learning the starting and ending position for that event in the context representation $H_X$. To do the same we use a span selector $\theta = [w^{(start)}, w^{(end)}] \in \mathbb{R}^{2h}$ with random initialization, as the learnable parameters. So, for an event feature $\lambda_k$ we have:
\begin{gather}
    \lambda^{(start)}_k = \lambda_k \circ w^{(start)} \in \mathbb{R}^h \\
    \lambda^{(end)}_k = \lambda_k \circ w^{(end)} \in \mathbb{R}^h
\end{gather}
where $\circ$ denotes element-wise multiplication. $\theta_k = [\lambda^{(start)}_k, \lambda^{(end)}_k]$ is the learnable span selector for $k^{th}$ slot in the prompt.

After getting the set of span selectors $\{\theta_k\}$ and the context representation $H_X$, for each $\theta_k$ we aim to extract a span $(\hat{st}_k, \hat{ed}_k)$ for the event $e_k$. For $\theta_k$ implying zero argument (when there is no argument for a particular event), the expected output $(\hat{st}_k, \hat{ed}_k) = (0, 0)$ which represents an invalid event category $\phi$. For the calculation of the distribution of each token being chosen as the start and end token, we do matrix multiplication ($\times$) of the span selector with the context $H_X$.
\begin{gather}
    logit^{(start)}_k = \lambda^{(start)}_k \times ({H_X})^T \in \mathbb{R}^{t_x} \\
    logit^{(end)}_k = \lambda^{(end)}_k \times ({H_X})^T \in \mathbb{R}^{t_x}
\end{gather}

where $logit^{(start)}_k$ and $logit^{(end)}_k$ represent the start and the end token distribution over the context tokens for $e_k$, and $t_x$ indicates the context token length, we calculate the probabilities where the start or the end positions are located by converting the distribution into a stochastic distribution using Softmax.
\begin{gather}
    p^{(start)}_k = SoftMax(logit^{(start)}_k) \in \mathbb{R}^{t_x} \\
    p^{(end)}_k = SoftMax(logit^{(end)}_k) \in \mathbb{R}^{t_x}
\end{gather}

We then define the loss function as the cross-entropy of the sum of $p^{(start)}_k$ and $p^{(end)}_k$.
\begin{gather}
    L_k(X) = -(\log p^{(start)}_k(st_k) + \log p^{(end)}_k(ed_k)) \\
    L = \sum_{X \in D} \sum_{k} L_k(X)
\end{gather}
 where $D \in [D_1, D_2, ... D_n]$ which makes up the entire dataset $\mathcal{D}$ and $k \in [1, l]$ denotes the event $e_k$ and $l$ is the number of event categories in prompt.

\subsubsection{Inference}
First, we define a set of spans for the events as $\mathcal{S} = \{(i, j) | (i, j) \in {t_x}^2, 0 < j - i \leq \alpha\} \cup \{(0, 0)\}$. It consists of all the spans that are less than the threshold $\alpha$ and a special span $\{(0, 0)\}$ to mark it as an invalid event. Our model extracts the event trigger related to the span selector $\theta_k$ by iterating and scoring all candidates as shown
\begin{gather}
    score_{(k)}(i, j) = logit^{(start)}_k(i) + logit^{(end)}_k(j) \\
    (\hat{st}_k, \hat{ed}_k) = \argmax_{(i, j) \in \mathcal{S}} score_{(k)}(i, j)
\end{gather}

where $(\hat{st}_k, \hat{ed}_k)$ is the predicted span for the event $e_k$ or slot $k$ in the prompt.

\section{Experiments}
\label{section:exp}

\begin{figure}[t!]
    \centering
    \includegraphics[scale=0.445]{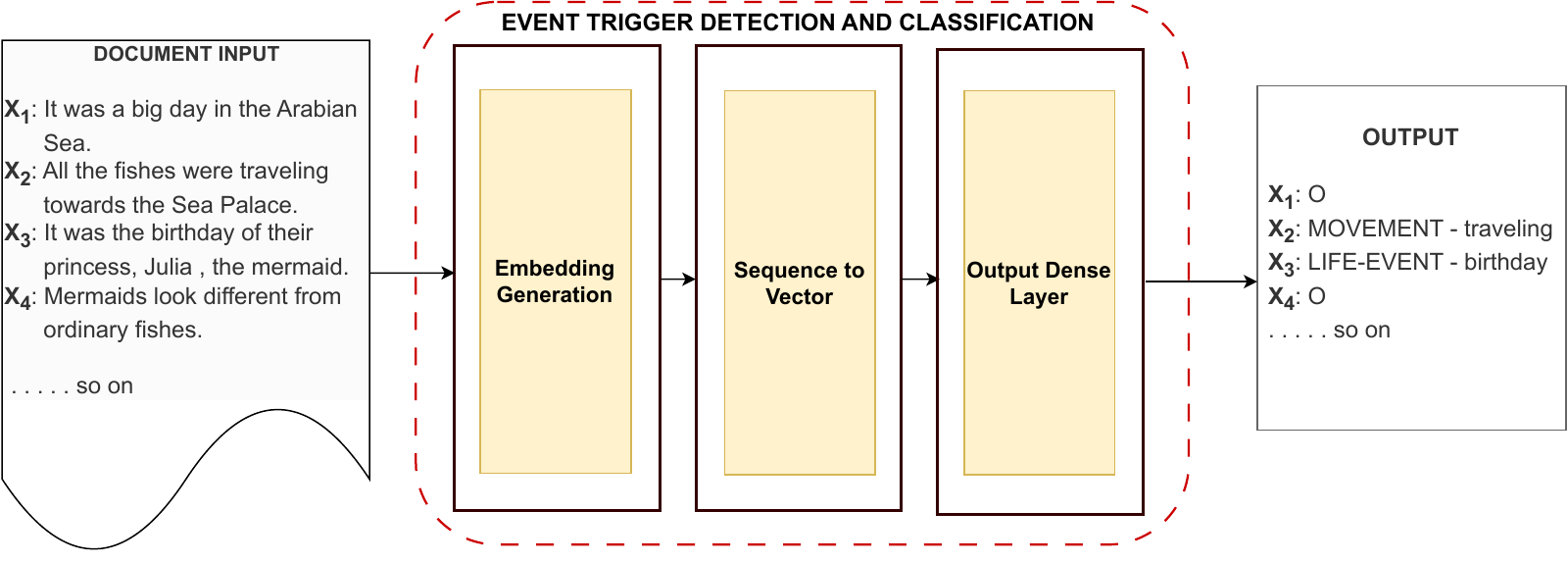}
    \caption{Block diagram illustrating the event detection framework.}
    \label{fig:overall}
\end{figure}

Previous research has shown that neural models are good at detecting events \cite{sims2019literary}. These models use the distributional information encoded in word embeddings. In this work, different variants of these models are analyzed, all of which take a sequence labeling approach, providing labels to tokens to indicate their event state. 
To achieve this, several neural models were explored in the literature. We chose to employ these models in our current work.
The task of event trigger detection is framed as a sequence labelling task. The overall framework is shown in \autoref{fig:overall}. The neural model in this framework consists of the following three modules -- (a) \emph{Embedding Generation}, (b) \emph{Sequence to Vector} and (c) \emph{Dense Layer}, which will learn the function
to predict one among the 7 output classes.

\subsection{Neural Methods}
\noindent \textbf{LSTM} 
This is the simplest model that has been considered as the baseline for the other models. The model that is under consideration is a single-direction, 100-dimensional LSTM, with each input token represented as a word embedding from Project Gutenberg.\\
\textbf{BiLSTM} The detection of a token as an event depends not only on the left context (context of the sentence that has come before the token) but also on the sentence on the right side (i.e., the tokens coming after the event has occurred). Consider the example below -  
\begin{quote} 
\centering 
``He \underline{came} from London." \\ 
``He \underline{came} down to fight." \\
\end{quote}
In the above statement, London plays an important role in classifying came as a separate type of event class, when compared with fighting. In the latter came describes a \texttt{CONFLICT} event class whereas in the first case, it was a \texttt{MOVEMENT} event class. Hence, it is clear that for better prediction we need forward as well as backward information.\\ 
\noindent\textbf{BiLSTM with Sentence CNN}
Several works have used a CNN on the sentence level for the task of event detection \citep{nguyen2015event}. We have employed the model used by \cite{sims2019literary} for event detection and modified the same for event classification.

\noindent\textbf{BiLSTM with Subword CNN} 
Subword character CNN captures meaningful representations of out-of-vocabulary words for learned embeddings. \\
\textbf{BiLSTM with Document Context} 
The main problem with BiLSTM is that the only information available is from the set, which may not always be enough to say exactly whether a given token represents an event. Therefore, to improve the performance of BiLSTM, we integrated the document context in addition to the sentence context already captured by the model. We have utilized the same model that \cite{sims2019literary} have used for this purpose. Necessary modifications were done for the task of event classification.

\noindent\textbf{BiLSTM with CRF} 
In this architecture, the BiLSTM is used to extract features from the input sequence, which are then used as inputs to the CRF~\citep{wang2018bidirectional}. 

\noindent\textbf{BERT fine-tuning} Fine-tuning BERT involves customizing a pre-trained BERT model for a specific task or domain by adjusting its parameters using a small amount of labelled data. For instance, in our scenario, fine-tuning involves using a dataset that includes texts paired with their corresponding event class labels for event classification with BERT. This involves adding a task-specific layer on top of the BERT encoder and training the entire model together, using a suitable loss function and optimizer.

\noindent\textbf{EVENT PROMPT} 

The training for the models was conducted on 120 short stories, with a validation set and testing set consisting of consisting of 20 and 60 short stories respectively. The optimizer that we have used is ADAM \citep{DBLP:journals/corr/KingmaB14} with Binary Cross-Entropy as the loss function for event trigger detection, and Categorical Cross-Entropy for predicting the categories of the event triggers detected. We have used ReLU \citep{agarap2018deep} as the activation function. The training of each model was done for 100 epochs. The batch size was 16, and it took about 2 hours to converge.

The above models primarily focused on trigger detection. We adapted these models for multi-class classification for the task of event classification.
In the field of event extraction, prompt-based generative approaches have recently emerged that offer greater flexibility than earlier classification-based methods.

We show the class-wise Precision, Recall, and F1 score for better comparison. In  \autoref{tab:hyperparameter_prompt}, we list the hyperparameters that have been used for training the model.

\begin{table}[]
\centering
\scalebox{0.55}{\resizebox{\textwidth}{!}{
\begin{tabular}{@{}lc@{}}
\toprule
\textbf{Hyperparameter} & \textbf{Values}          \\ \midrule
BART Output Dimension   & 768                    \\
Epoch                   & 1000                    \\
Batch size              & 8                        \\
Optimizer               & AdamW                    \\
Scheduler               & Linear (with 0.1 warmup) \\
Max span length         & 10                       \\
Max gradient norm       & 5                        \\
Max encoder seq length  & 500                      \\
Max decoder seq length  & 80                       \\
Weight Decay            & 0.01                     \\
Loss function           & Cross-Entropy            \\
Seed                    & 42                       \\
Learning Rate           & 4e-5                     \\ \bottomrule
\end{tabular}}}
\caption{Hyperparameters and their values that are used for training the model.}
\label{tab:hyperparameter_prompt}
\end{table}

\section{Results and Discussion}
\label{section:res}

The event trigger detection results obtained from the baseline models and prompt-based model are shown in~\autoref{tab:event_trigger}. 
It is observed that the proposed model \textbf{EVENT PROMPT} yields an F1 score of 89.9\% which is higher than the baseline models. 
In a previous study, from which we adopted baseline models~\citep{sims2019literary}, the F1 scores achieved for various models utilized in event detection from literary texts were lower than those obtained for short stories.
In the case of event classification, it is observed that except for the class {\tt{MOVEMENT}}, the other classes classified by \textbf{EVENT PROMPT} with an F1 score are better than the other baseline models. Out of all the classes, the {\tt{COMMUNICATION}} class yields the highest F1 score, and the class {\tt{OTHERS}} have the lowest. This might be due to the greater number of events for the class {\tt{COMMUNUCATION}} as compared to the other classes as shown in~\autoref{tab:event_typ}. We noted a significant imbalance in the distribution of event types within the dataset.

\begin{table}[t!]
\centering
\scalebox{0.65}{\resizebox{\textwidth}{!}{
\begin{tabular}{@{}lcccccc@{}}
\toprule
{\textbf{METHOD}} 
                                 & \textbf{PRECISION} & \textbf{RECALL} & \textbf{F1}    \\ \midrule
\textbf{EVENT PROMPT}       & 85.35 & 87.23 & \textbf{86.28} \\
\bottomrule
\end{tabular}%
}
}
\caption{Result of Event trigger detection}
\label{tab:event_trigger}
\end{table}

\begin{figure}[t]
    \centering
    \includegraphics[scale=0.59]{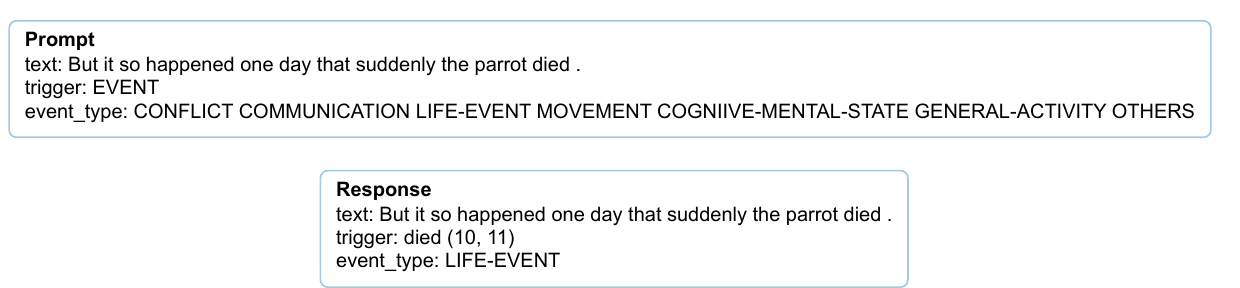}
    \caption{Contextualized prompt provided to the model to elicit both the trigger and event type.}
    \label{fig:promptoutput}
\end{figure}

\begin{table}[]
\centering
\scalebox{0.90}{\resizebox{\textwidth}{!}{%
\begin{tabular}{@{}lcccccccc@{}}
\toprule
\textbf{METHOD} &
  \textbf{SCORE} &
  \textbf{CON} &
  \textbf{COM} &
  \textbf{\begin{tabular}[c]{@{}c@{}}LE\end{tabular}} &
  \textbf{MOV} &
  \textbf{\begin{tabular}[c]{@{}c@{}}CMS\end{tabular}} &
  \textbf{\begin{tabular}[c]{@{}c@{}}GA\end{tabular}} &
  \textbf{OTH} \\ \midrule
                                                 & P  & 87.5 & 90.8   & 69.4 & 73.3 & 77.5 & 68.1 & 31.3 \\
                                                  & R  & 46.2  & 91.8 & 47.9 & 79.3 & 76.7 & 64.8 & 16.3 \\
\multirow{-3}{*}{\textbf{LSTM}}                      & F1 & 60.5 & 91.3 & 56.6 & 76.2 & 77.0 & 66.4   & 21.4 \\ \midrule
                                                     & P  & 70.6 & 91.7 & 72.2 & 75.2 & 75.2 & 68.7 & 38.1 \\
                                                     & R  & 42.9 & 90.9 & 43.4 & 81.0 & 76.3 & 69.7 & 18.5 \\
\multirow{-3}{*}{\textbf{BiLSTM}}                    & F1 & 53.1 & 91.3 & 54.2 & 78.0 & 75.7 & 69.2 & 24.9 \\ \midrule
                                                     & P  & 0.0    & 91.5 & 81.1 & 78.9 & 82.3 & 75.8 & 25.6 \\
                                                     & R  & 0.0    & 85.6 & 21.1 & 65.0   & 47.9 & 41.3 & 5.6 \\
\multirow{-3}{*}{\textbf{~~+Document Context}}         & F1 & 0.0    & 88.4 & 33.3 & 71.3 & 60.6 & 53.4 & 9.1   \\ \midrule
                                                     & P  & 69.2    & 91.9 & 61.6 & 76.9   & 77.9 & 69.0 & 40.4 \\
                                                     & R  & 32.0    & 91.3 & 55.9 & 80.0   & 73.6 & 72.8 & 18.5 \\
\multirow{-3}{*}{\textbf{~~+Sentence CNN}}             & F1 & 43.8    & 91.6   & 58.6 & 78.4 & 75.7 & 70.8 & 25.4 \\ \midrule
                                                     & P  & 75.0 & 91.9 & 65.0 & 80.0 & 80.0 & 71.8 & 35.5   \\
                                                     & R  & 53.6 & 92.8 & 48.4 & 80.5 & 77.1 & 71.5 & 30.7 \\
\multirow{-3}{*}{\textbf{~~+Char CNN}}                 & F1 & 62.3 & 92.3 & 55.5 & 80.2 & 78.5 & 71.6 & 32.9 \\ \midrule
                                                     & P  & 93.7 & 90.0 & 66.6 & 83.5   & 79.6 & 72.4 & 47.3 \\
                                                     & R  & 50.0 & 95.5 & 60.0 & 86.0 & 81.4 & 80.0 & 36.8 \\
\multirow{-3}{*}{\textbf{~~+CRF}}                     & F1 & 65.2 & 92.7 & 63.1 & 84.7 & 80.5 & 76.0 & 41.6   \\ \midrule
                                                     & P  & 90.9 & 91.1 & 70.7 & 83.2   & 80.7 & 78.6 & 41.2 \\
                                                     & R  & 32.0 & 95.7 & 65.8 & 91.2 & 85.6 & 72.3 & 44.5 \\
\multirow{-3}{*}{\textbf{BERT (fine-tuned)}}                     & F1 & 47.1 & 93.4 & 68.2 & 87.0 & 83.0 & 75.3 & 42.8   \\ \bottomrule
                                                    & P  & 77.3 & 93.7 & 72.2 & 84.1 & 84.3   & 75.3 & 48.8 \\
                                                     & R  & 62.9  & 93.4 & 70.5 & 81.5 & 84.0 & 80.1 & 43.8 \\
\multirow{-3}{*}{\textbf{EVENT PROMPT}}             & F1 & \textbf{69.4} & \textbf{93.6} & \textbf{71.4} & 82.8   & \textbf{84.2} & \textbf{77.6} & \textbf{44.3} \\
                                                     \bottomrule
\end{tabular}%
}
}
\caption{ Overall performance of event classification. The best result is highlighted.}
\label{tab:event classification}
\end{table}

\begin{figure}[ht]
    \centering
    \includegraphics[scale=0.5]{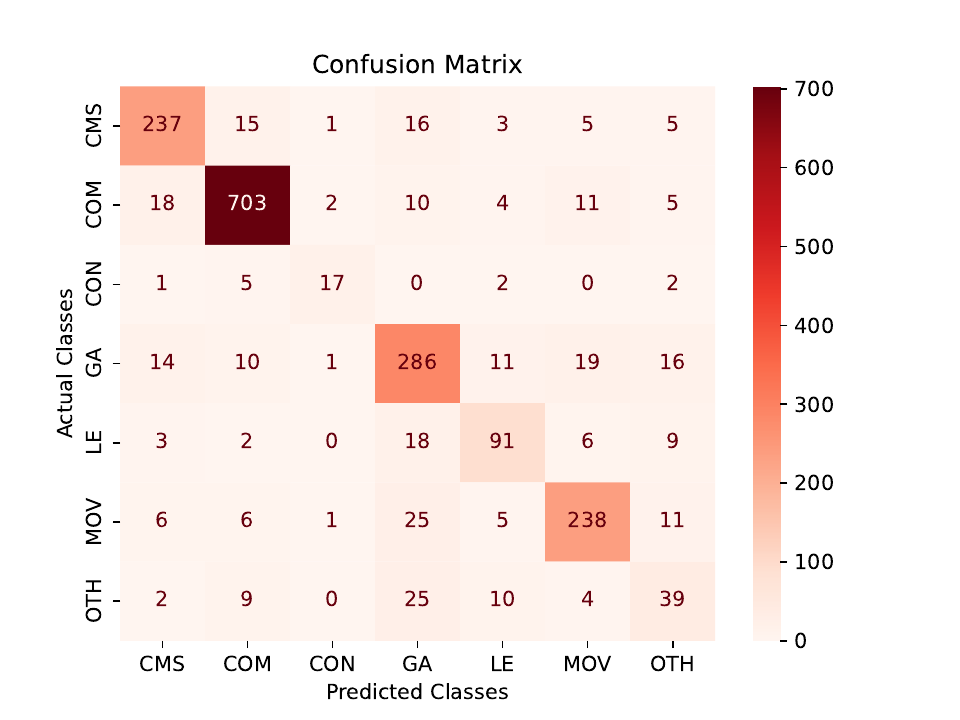}
    \caption{Confusion matrix of our proposed model}
    \label{fig:conf}
\end{figure}

This factor can be attributed to the fact that in children's short stories, the teachings, morals, discipline, etc., taught to the children are narrated through verbal communication between the subjects of the stories. The target audience for the short stories is children, and the aim of these stories is to nurture young minds with positive and constructive ideas while steering them away from harmful or negative thoughts.
This fact is clearly demonstrated by the data presented in Table \ref{tab:event_typ}. It shows a notable disparity in the distribution of event types, with the number of events classified as {\tt{CONFLICT}} being significantly lower compared to the maximum number of events falling under the {\tt{COMMUNICATION}} category. \autoref{tab:ann_data} clearly illustrates that most events depicted in children's literature are based on reality. These narratives are grounded in real-life experiences, facts, and occurrences, emphasizing practical teachings and tangible lessons. Conversely, instances of purely imaginative or fantastical ideas are comparatively scarce within this genre. This indicates a deliberate emphasis on imparting realistic knowledge and values to young readers, enriching their understanding of the world around them.
\autoref{fig:promptoutput} illustrate an example demonstrating the detailed response provided by our proposed model. The input prompt, as well as the output from the model, are shown. 
In \autoref{tab:event classification}, we can observe that the best-performing model from the previously discussed methods is outperformed by the prompt-based approach, especially for the categories with very few occurrences (e.g. {\tt{CONFLICT}} class). This is mainly because of the presence of better context features before the extraction of the span. 
Better analysis can be done by observing the confusion matrix shown in~\autoref{fig:conf} for the \textbf{EVENT PROMPT} model. The maximum misclassification can be seen in {\tt{GENERAL-ACTIVITY}} and {\tt{OTHERS}} class. Mostly, the class {\tt{OTHERS}} was misclassified from the other classes as the scope of this class is greater compared to the other classes. The annotators get confused when marking this event class.

\subsection{Data Augmentation}

The gold standard data is necessary to create any Natural Language Processing (NLP) application. The annotation process has become more difficult as NLP has attempted to handle increasingly complex linguistic phenomena or a variety of labelling and classification tasks, yet the requirement for and usefulness of large labelled datasets persist~\citep{sun2017revisiting, human-annotation}. Manual annotating linguistic resources for training and testing is expensive, especially for specialized text fields. For the same reason, we tried to generate the labels for event extraction and classification automatically from \textbf{EVENT PROMPT}. 
The approach incorporates the following steps:
\begin{enumerate}
    \item Manually annotate 200 short stories by a single human expert, which we can use for further processing. 
    \item Train \textbf{EVENT PROMPT} models on the manually annotated data.
    \item Utilize \textbf{EVENT PROMPT} model to synthetically generate labels for the 800 unlabelled short stories, which can be further enhanced by utilizing human expertise.
    \item Benchmark the final dataset of 1000 and compare the performance.
\end{enumerate}

The validity of the expanded dataset is performed in two sets of experiments:
\begin{itemize}
    \item We randomly selected 120 short stories from the synthetically expanded data and trained the \textbf{EVENT PROMPT} model keeping the dev and test set same as earlier.
    \item To see the effect of large training data, we selected 500 synthetically generated data and trained the model with this data keeping the dev and test set same as earlier. 
\end{itemize}

The experiments' results for event trigger detection and event classification are presented in~\autoref{tab:synthetic_trigger} and~\autoref{tab:synthetic_classification}. It was observed that the performance degraded slightly when trained with 120 synthetically labelled short stories. However, training the model with 500 synthetically labelled short stories resulted in improved performance, outperforming some classes in the task of event classification.

\begin{table}[]
\centering
\scalebox{0.70}{\resizebox{\textwidth}{!}{%
\begin{tabular}{@{}lccc@{}}
\toprule
\textbf{MODEL}       & PRECISION & RECALL & F1   \\ \midrule
\textbf{EVENT PROMPT}        & 87.5      & 92.4     & 89.9 \\
\textbf{Trained on 120 (synthetically generated)} & 83.2     & 87.1   & 85.1 \\
\textbf{Trained on 500 (synthetically generated)}         & 86.6      & 89.7   & 88.1 \\ \bottomrule
\end{tabular}}}
\caption{Comparison of the \textbf{EVENT PROMPT} model for different dataset curation methods for event trigger detection.}
\label{tab:synthetic_trigger}
\end{table}

\begin{table}[]
\centering
\resizebox{\textwidth}{!}{%
\begin{tabular}{@{}lcccccccc@{}}
\toprule
\textbf{Model} &
  \textbf{SCORE} &
  \textbf{CON} &
  \textbf{COM} &
  \textbf{\begin{tabular}[c]{@{}c@{}}LE\end{tabular}} &
  \textbf{MOV} &
  \textbf{\begin{tabular}[c]{@{}c@{}}CMS\end{tabular}} &
  \textbf{\begin{tabular}[c]{@{}c@{}}GA\end{tabular}} &
  \textbf{OTH} \\ \midrule
\multirow{3}{*}                                                            & P  & 77.3 & 93.7 & 72.2 & 84.1 & 84.3   & 75.3 & 48.8 \\
                                                     & R  & 62.9  & 93.4 & 70.5 & 81.5 & 84.0 & 80.1 & 43.8 \\
\multirow{-3}{*}{\textbf{EVENT PROMPT}}             & F1 & 69.4 & 93.6 & 71.4 & 82.8   & 84.2 & 77.6 & 44.3 \\ \midrule
\multirow{3}{*}{\textbf{Trained on 120 (synthetically generated)}} & P  & 63.7 & 87.1 & 79.3 & 80.2 & 72.1 & 68.1 & 15.4 \\
                                        & R  & 54.9 & 96.5 & 43.2 & 78.8 & 81.7 & 66.5 & 61.4 \\
                                        & F1 & 58.9 & 91.6 & 55.9 & 79.4 & 76.6 & 67.3 & 24.6 \\ \midrule
\multirow{3}{*}{\textbf{Trained on 500 (synthetically generated)}}         & P  & 68.4 & 91.3 & 79.4 & 81.8 & 86.8 & 75.1 & 50.9 \\
                                        & R  & 58.7 & 93.4 & 63.1 & 90.9   & 75.7 & 83.4 & 47.3   \\
                                        & F1 & 63.1 & 92.3 & 70.3 & 86.1 & 80.8 & 79.0   & 49.0   \\ \bottomrule
\end{tabular}%
}
\caption{Comparison of the best-performing model for different dataset curation methods for event classification.}
\label{tab:synthetic_classification}
\end{table}

\subsection{Ablation Study}
\label{section:ablation}
\begin{enumerate}
\item \textbf{Effect of Learning Rate:}
 We conducted an ablation study to evaluate the impact of various learning rates on the performance of the \textbf{EVENT PROMPT} model. The results, presented in~\autoref{tab:lr}, indicate that the model achieved its highest performance with a learning rate of 4e-5. At this optimal learning rate, the model attained a macro F1 score of 74.74\%, highlighting the sensitivity of the model's performance to changes in the learning rate parameter. This finding underscores the importance of carefully tuning hyperparameters to enhance model effectiveness.
\item \textbf{Training with 120 synthetic short stories:} Training the \textbf{EVENT PROMPT} model with 120 synthetic short stories resulted in degraded performance. The impact of using Silver standard data is clearly reflected in the model's performance. This degradation suggests that the quality of the training data plays a crucial role in the effectiveness of the model. High-quality, authentic data may be necessary to achieve optimal results, as synthetic or lower-quality data can negatively affect the model's ability to generalize and perform well on real-world tasks.
    
\item \textbf{Training with 500 synthetic short stories:} Furthermore, the \textbf{EVENT PROMPT} model was trained with 500 synthetically labelled short stories. The performance results, shown in \autoref{tab:synthetic_trigger} and \autoref{tab:synthetic_classification}, indicate that the degradation in performance caused by the Silver standard data can be mitigated by increasing the volume of such data. This suggests that while Silver standard data may initially lead to a decrease in model performance, a larger quantity of this data can help compensate for its lower quality, ultimately enhancing the model's overall effectiveness. This finding underscores the potential of using extensive synthetically labeled datasets to improve model performance when high-quality labeled data is limited.
    
\end{enumerate}

\begin{table}[]
\centering
\scalebox{0.46}{\resizebox{\textwidth}{!}{%
\begin{tabular}{@{}ccc@{}}
\toprule
\textbf{Learning Rate}       & \textbf{F1 (macro) in\%}    \\ \midrule
1e-5        & 71.23       \\
2e-5        & 73.67       \\
3e-5         & 71.94       \\
4e-5         & 74.74       \\
5e-5         & 74.59       \\
6e-5         & 74.17       \\
\bottomrule
\end{tabular}}}
\caption{F1-Score (macro) for different learning rate for the event classification task}
\label{tab:lr}
\end{table}

\section{Error Analysis}
\label{section:error}
This section describes the errors encountered while predicting realis events using the best-performing \textbf{EVENT PROMPT} model. The errors can be categorized into two main types based on the subtasks: event detection and event classification.

For the task of event detection, it was observed that most errors stemmed from incorrectly classifying unreal events as real ones. This might be due to subtle contextual cues or ambiguities in the training data. Although the performance was fairly good, the thin line between realis and non-realis events sometimes creates confusion. Improving the model's ability to detect these nuances may require more sophisticated feature extraction or additional training on more diverse and representative datasets.

In the context of event classification, the class with the least performance is {\tt{OTHERS}}. This is due to the presence of an almost negligible amount of this type of event in the dataset. The model didn't have enough samples to train on. Additionally, the misclassification between the {\tt{GENERAL-ACTIVITY}} and {\tt{OTHERS}} categories was higher because the {\tt{GENERAL-ACTIVITY}} class has the widest scope of events, while the {\tt{OTHERS}} category acts as an outlier class.


\section{Conclusions}
\label{section:conc}

In this work, we proposed a prompt-based approach for event trigger detection and classification in short stories using contextualized prompts. We first compared traditional feature extraction models with the prompt-based approaches. We tried to address problems faced when working with event extraction, like the unavailability of good-quality annotated datasets and data imbalances.
We have used a pre-trained language model built on the standard transformer, BART.
The results show that the proposed method outperforms the baseline models. Since the target audience for short stories is children, we observed an imbalance in the number of events in the dataset (e.g. less number of events for {\tt{CONFLICT}} class and more for {\tt{COMMUNICATION}} class).
It was also observed that the heavily imbalanced classes improved through contextualized prompts. 
We got a performance improvement from \textbf{65.2 to 69.4\%} in F1-score for \texttt{CONFLICT} category, which had a significantly less number of instances throughout the dataset. 
The best-performing model \textbf{EVENT PROMPT} was used to expand the dataset and annotate 800 stories synthetically, making a complete dataset of 1000 short stories. This synthetically generated dataset was validated with extensive experiments.
Prompt-based methods are, therefore, found to be efficient for small and imbalanced datasets. Overall, contextualized prompts are a powerful tool for enhancing the performance of event extraction models by providing them with the necessary context and guidance to extract events accurately from text and classify them.
The domain of Prompt-Based learning is growing and is flexible enough to be used in most information retrieval tasks. There are a lot of scopes available in argument mining and other subtopics. 


\bibliography{template}

\end{document}